# Segregation competition and complexion coexistence within a polycrystalline grain boundary network

Pulkit Garg [a,1], Zhiliang Pan [a,b,1], Vladyslav Turlo [a,c], Timothy J. Rupert [a,*]

[a] Department of Materials Science and Engineering, University of California, Irvine, CA 92697, USA

[b] School of Mechanical and Electrical Engineering, Guilin University of Electronic Technology Guilin, Guangxi Province 541004, China

[c] Laboratory for Advanced Materials Processing (LAMP), Swiss Federal Laboratories for Materials Science and Technology (Empa), Thun, CH-3602, Switzerland

[1] These authors contributed equally to this work.
* To whom correspondence should be addressed: trupert@uci.edu

Interfacial segregation can stabilize grain structures and even lead to grain boundary complexion transitions. However, understanding of the complexity of such phenomena in polycrystalline materials is limited, as most studies focus on bicrystal geometries. In this work, we investigate interfacial segregation and subsequent complexion transitions in polycrystalline Cu-Zr alloys using hybrid Monte Carlo/molecular dynamics simulations. No significant change in the grain size or structure is observed upon Zr dopant addition to a pure Cu polycrystal at moderate temperature, where grain boundary segregation is the dominant behavior. Segregation within the boundary network is inhomogeneous, with some boundaries having local concentrations that are an order of magnitude larger than the global value and others having almost no segregation, and changes to physical parameters such as boundary free volume and energy are found to correlate with dopant concentration. Further, another alloy sample is investigated at a higher temperature to probe the occurrence of widespread transitions in interfacial structure, where a significant fraction of the originally ordered boundaries transition to amorphous complexions, demonstrating the coexistence of multiple complexion types, each with their own distribution of boundary chemical composition. Overall, this work highlights that interfacial segregation and complexion structure can be diverse in a polycrystalline network. The findings shown here complement existing computational and experimental studies of individual interfaces and help pave the way for unraveling the complexity of interfacial structure in realistic microstructures.

# 1. Introduction

In nanocrystalline metals, grain boundaries (GBs) act as nucleation and pinning sites for dislocations, contributing to the greatly elevated strengths of these materials [1-4]. However, grain growth can be a major problem since GBs have higher energies than the bulk crystalline region, making many nanocrystalline metals thermally unstable and causing them to lose their advantage of high strength through rampant coarsening [5-7]. To retain the superior properties of nanocrystalline metals, solute addition is commonly used to restrict boundary migration and/or reduce the energy penalty for having a GB, thereby stabilizing the microstructure [8-20]. While solute segregation is desired for nanocrystalline alloys, the extent of segregation can vary for different GBs [21, 22]. Therefore, an understanding of how segregation is dependent on the GB structure and the variety of GB concentrations expected for a polycrystal is crucial for realizing tunable alloy microstructure and properties of nanocrystalline alloys.

An early theory for GB segregation developed by McLean [23] identified elastic strain energy from the degree of solute misfit in a solution as the critical variable and then determined the solute atomic fraction at a monolayer interface using statistical mechanics. Subsequently, several refinements to this type of model have been made to include the effects of chemical contribution from solute segregation [24], structural and energy anisotropy of GBs [25], and interphase boundaries in multicomponent alloys [26]. In addition, the dependence of GB segregation on misorientation angle has been extensively investigated through experiments and theoretical modeling [27-30]. In the Fe-Si system, Watanabe et al. [31] showed that the amount of Si segregation increased with increasing misorientation angle of tilt GBs, but no prominent trend was observed for the twist boundaries. Such a lack of correlation was also observed for different GBs in another Fe-Si system [32] and a Fe-C system [33], as well as in symmetric tilt GBs in Cu-Sb bicrystals [27]. Dingreville and Berbenni [28] used a continuum linear defect mechanics model to study GB segregation, finding that the susceptibility for H segregation first decreased and then increased with increasing misorientation angle of <001> symmetric tilt GBs in Ni. In contrast, the opposite trend for H segregation at different GBs in Ni was demonstrated by Jothi et al. [30] using a finite element analysis-based numerical modeling framework. On the other hand, the segregation of Au to GBs in Pt was found to increase with increasing GB energy [34]. While GBs have complex structures, some of the features describing GB character appear to be more universal descriptors and are correlated to each other [35-37]. According to the Read-Shockley relation [38], GB energy increases with increasing misorientation angle for low-angle boundaries and can be calculated indirectly from the experimental measurement of misorientation angle [39]. Despite this mounting evidence of variety in GB behavior, the vast majority of existing models assume that all GBs in a polycrystal have the same tendency for solute segregation. This assumption simplifies the models significantly but is inaccurate [40], as very recent work has demonstrated that a broad spectrum of possible atomic segregation sites are available within polycrystals [41, 42]. Further work is needed to uncover the variety of solute segregation behavior that is possible and understand how different GB parameters affect this phenomenon in polycrystalline materials.

Quantitative experimental analysis of GB segregation is extremely challenging since each of the five crystallographic parameters that describe an interface can conceivably affect segregation. Transmission electron microscopy is commonly used to report on GB segregation (see, e.g., [43, 44]), a novel aberration-corrected techniques are now able to enable composition measurements with atomic resolution [45, 46]. Unfortunately, these studies require all measured GBs to be in an edge-on condition, which is particularly difficult to achieve for nanocrystalline alloys where the grains are small and overlapping grains can regularly occur [47]. Perhaps more problematic is the fact that the electron beam used to extract a composition measurement has an interaction volume that is typically larger than the GB thickness, meaning that any measurement of "GB concentration" is really a value averaging among the GB itself plus some portion of the neighboring grains. As a result, many experimental studies of segregation instead report on GB excess, or the number of atoms per unit cross-sectional GB area. For example, Hu et al. used GB excess while discussing how the orientation of the lower-Miller index grain surface controls the resultant interfacial structure in Ni-S [48]. While this metric is more reliably obtained, as local area scans containing the GB can be compared to regions in the crystal, it does not account for variations in GB thickness, which can be important for comparing boundaries with different character and especially if comparing boundaries which have undergone structural transitions. Further, many experimental segregation studies have focused on bicrystal samples with special configurations [24, 27, 28] which, while intriguing, provide a limited view of segregation phenomena. Perhaps the most robust method of studying segregation is the combined use of atom probe tomography and transmission electron microscopy to provide reliable measurements of GB concentration and complementary imaging of GB structure, with the work of Herbig et al. on C segregation at GBs in cold-drawn pearlitic steel being an instructive example [33]. However, such studies are time and resource intensive, while also usually providing limited statistics as only a limited population of GBs can be studied in most cases [49].

While atomistic simulations offer the highest resolution for characterizing GB segregation, these simulations are often limited to bicrystal geometries [29, 30, 37]. First-principles simulations can provide insight into the atomic bonding between solute and matrix atoms, but are restricted to highly symmetric GBs with small simulation cells and also typically ignore the effect of temperature on segregation [50, 51]. Olmsted et al. [52] examined a large number of GBs using bicrystal samples with molecular dynamics (MD) but no strong correlations were observed between the GB energy and boundary properties such as disorientation angle and free volume. Furthermore, the presence of metastable states in addition to the minimum energy structures makes it important to study an ensemble of GBs in a polycrystalline sample [53]. Recent work by Aksoy et al. [54] showed that the embrittlement potency of an isolated symmetric tilt GB depends on its local energy but, in a polycrystal, this quantity also depends on the probability of solute segregation level at the specific GB. Thus, GB segregation studies in bicrystal geometries are limited by their use of isolated and symmetric interfacial structures which may not be representative of the segregation behavior in a full polycrystalline system.

Solute segregation can also induce interfacial structural transitions in polycrystalline alloys, leading to the formation of solute-rich interfacial phases or *complexions* [55-57]. All interfaces in a polycrystalline material do not transition simultaneously, with complexion transitions characterized by a discontinuous change in thermodynamic parameters such as excess entropy, excess volume, and the first derivative of the excess energy of the GB [58]. Measuring a discontinuous change in thermodynamic properties of a boundary is limited by several assumptions and requires the measurement of many interfaces in a polycrystalline material [59, 60] making it very difficult to directly observe complexion transitions. Measurable features such as GB thickness, mobility, structure motifs, and solute concentration are often used to identify different types of complexions [61-63]. The availability of experimental techniques for accurate measurements of local GB thickness has made this metric a common criterion for categorizing complexions [64]. Six different complexion types based on thickness were proposed by Dillon et al., including monolayer solute segregation, clean GBs, bilayer solute segregation, multilayer segregation, nanoscale intergranular films, and wetting films [61]. While not the explicit focus of this classification scheme, the extent of structural disorder generally increases with increasing complexion thickness. Pan and Rupert [65] showed that the GB segregation trends and the formation of amorphous complexions are determined by the GB core structure. Expanding on such a concept, Hu et al. [66] used genetic algorithms to develop a neural network model for GB segregation for a large number of bicrystal GBs as temperature and global composition were varied. While these authors were primarily focused on measuring GB composition and creating GB phase diagrams, local structural disorder and local free volume were also investigated and predicted, demonstrating that boundary structure is also an important variable to consider. Specifically, we hypothesize here that multiple complexions types will often coexist and GB descriptors will widely vary within a polycrystalline GB network, necessitating a further understanding of segregation competition and the distribution of complexion structures within a single material.

In this work, atomistic simulations are used to investigate the atomic-scale details of GB segregation and segregation-induced complexion transitions in polycrystalline Cu-rich alloys. Hybrid Monte Carlo (MC)/MD simulations are used to create equilibrated samples of pure Cu and Cu-Zr polycrystals at two different temperatures. At a moderate temperature of 900 K, solute segregation at the GB network is inhomogeneous, allowing for conclusions about segregation distributions to be made, yet no significant change in the grain structure is observed as compared to the pure Cu polycrystal, allowing for a comparison of each segregated state with its starting boundary structure. Alterations to the starting GB free volume and GB energy in the doped Cu-Zr polycrystal at 900 K are found to correlate strongly with GB solute concentration. In this sample, a vast majority of the GBs remain ordered and only a few boundaries appear to transform into thin amorphous complexions. To promote more widespread complexion transitions, a second Cu-Zr polycrystal is investigated at 1050 K, where a significant number of boundaries transition to amorphous complexions with varying thicknesses. Even within a given amorphous complexion, significant chemical heterogeneity is observed. One benefit of a fully resolved atomistic study of the GB network is that all complexion types and thicknesses can be captured, allowing for metrics such as the transformed fraction to be extracted. Since this metric is difficult or impossible to obtain from experiments, we demonstrate how polycrystalline models can be compared with existing experimental data for amorphous complexion populations to aid their interpretation. Overall, the resolution offered by atomistic simulations makes it possible to fully characterize GB segregation and complexion coexistence in a polycrystalline system, with large variations in the interfacial behavior observed.

## 2. Computational methods

Hybrid atomistic MC/MD simulations were performed with the Large-scale Atomic/Molecular Massively Parallel Simulator (LAMMPS) code [67], using an integration time step of 1 fs for all MD simulations. Cu-Zr was chosen as a representative alloy system since Zr readily segregates at GBs due to its low solubility in Cu [68], and also complexion transitions from ordered to amorphous boundary structure have been observed in this system by both simulations [65] and experiments [19, 57]. The tendency for forming amorphous complexions at high temperature can be tied to the fact that Cu-Zr is a good glass forming alloy system [69]. Interactions between Cu-Cu and Zr-Zr atoms were described with embedded-atom method formulations, while interactions between Cu and Zr atoms were defined using a Finnis-Sinclair formulation [70].

First, a pure Cu polycrystalline sample was constructed using a Voronoi tessellation method in a $37\times37\times37$ nm$^3$ cubic box, with periodic boundary conditions applied in all three directions. The starting sample contains 100 grains, with an average grain size of 10.1 nm and ~4,000,000 atoms. The system was first equilibrated with a conjugate gradient minimization technique and then further relaxed with a Nose-Hoover thermo/barostat for 200 ps under zero pressure at 900 K. During this annealing step, the grain structure relaxes, and the starting grain boundaries change local structure to lower the system energy. To understand the segregation behavior at the GB network, the polycrystalline sample was then doped with Zr atoms at 900 K using the hybrid MC/MD method with MC steps performed in a variance-constrained semi-grand canonical ensemble [71]. MC switches occur after every 100 MD steps, with the global composition fixed to 2 at.% Zr by adjusting the chemical potential difference during the simulation [72]. To facilitate a greater number of transitions of ordered GBs to amorphous complexions, the alloy was also simulated at a higher temperature of 1050 K (~0.8 melting temperature of pure Cu (1353 K) for this potential) with the same hybrid MC/MD approach. The simulations were determined to have converged to an equilibrated state if the gradient of the system's potential energy over the last 4000 MC steps is less than 0.001 eV/step [65]. At this point, the energy of the system has reached a plateau and is no longer decreasing noticeably, with only small fluctuations around a constant value observed at this point. No changes in the grain structure or dopant distribution are observed with further simulation steps. Simulation convergence was very challenging, with the run time of each taking more than two months on nine computing nodes containing 216 processors and costing approximately 0.4M CPU hours each.

After the equilibrium structure of both the clean and doped polycrystal is reached, another conjugate gradient minimization was used to remove thermal noise while preserving the defect structures obtained. Atomic configurations were then analyzed with adaptive common neighbor analysis (CNA) [73, 74], which isolates the local structural environment of an atom. The Grain

Tracking Algorithm (GTA) code [75-77] enabled the identification of individual grains based on a comparison of the local crystallographic orientation of each atom and characterization of each GB in terms of macroscopic metrics such as misorientation angle and boundary plane normal. Due to the symmetry of the face centered cubic (FCC) crystal structure, the misorientation angle between neighboring grains can have multiple different values depending on the choice of the rotation axis. In this work, the smallest misorientation angle was used, henceforth referred to as the *disorientation angle*, to indicate the misorientation between the two grains. Neighboring crystal atoms with disorientations of less than 1° are identified as belonging to the same grain. GBs were then identified as intergranular defects with two neighboring grains using the GTA code. 370-470 interfaces were identified and studied for each sample, giving a large number of GBs from which statistical trends in segregation and complexion transitions behavior could be identified. Triple junctions were identified as separate defects, using the fact that they have three neighboring grains during analysis with the GTA, and were not included in the analysis presented here to keep the focus on grain boundary complexions.

To capture the physical environment of each GB atom, the excess atomic energy and excess atomic volume were also calculated and compared to the grain interior atoms, giving measurements of GB energy and GB free volume, respectively. We note that this definition of GB energy is most accurately described as the excess potential energy of the GB atoms and does not include contributions such as entropy. Thermodynamic models of interfacial transitions often require more nuanced treatments including a comparison versus a bulk region with similar composition and the contribution from vibrational entropy (see, e.g., [78]). To track the evolution of the GB network due to the doping process, grain number was mapped based on the coordinates of the innermost grain atoms to ensure that each grain has the same identification number before and after doping. A GB that has the same neighboring grains before and after doping is therefore identified as the same defect. Each GB can be assigned an average Zr concentration, defined as the ratio of the number of Zr atoms to the total number of atoms in the GB. The thickness of a GB is measured by fitting two parallel planes, one on each side of the GB viewed in an edge-on condition. The largest plane spacing which includes only GB atoms, without any grain atoms, corresponds to the thickness of a given boundary. This is equivalent to measuring the minimum thickness of a given complexion, matching experimental procedures used in the literature [47, 79]. All atomic configurations were visualized using the open-source visualization tool OVITO [80].

## 3. Results and discussion
### 3.1 GB segregation trends

To begin to probe the simplest case of polycrystalline segregation, we first analyze the nanocrystalline grain structure in pure Cu equilibrated at 900 K (Figure 1(a)) and the Cu-Zr alloy equilibrated at 900 K (Figure 1(b)). Minimal grain growth was observed upon equilibrating either the pure (mean grain size of 10.1 nm) or alloyed sample (mean grain size of 10.2 nm) at a temperature of 900 K, suggesting little change from the original grain structure. In this figure, Cu atoms are colored according to their local crystal structure identified with CNA, with FCC atoms colored green, hexagonal close packed (HCP) atoms colored red, other or non-crystalline atoms colored white, while Zr atoms are colored black. It is clear from Figure 1(b) that the vast majority of Zr atoms are found to segregate to the GB, with only a few isolated Zr atoms observed in the grain interiors. The mean composition of the grain interior is 0.1 at.% Zr, slightly below the solubility limit of Zr in FCC Cu [57], while the mean GB composition is 9.4 at.% Zr (Note: the statistical description of the GB concentration distribution will be studied in detail shortly), supporting the visual observations. Solute segregation at the GB network should be closely related to the GB crystallography and therefore also the local atomic structure of interfaces prior to doping, as these will dictate possible sites for segregation and the density of such sites. The structure and energy of specific GBs in the pure sample can serve as reference states for the same boundaries in the doped sample, provided the GB network has not evolved between the two simulations. While the similarity in grain size between the pure and alloyed samples is a promising signal, it is important to first determine more rigorously if there have been changes to the nanocrystalline grain structure upon solute addition before proceeding to further analysis. To this end, the disorientation angles between neighboring grains in polycrystalline pure Cu and the Cu-Zr alloy were extracted and compared.

Figure 2(a) shows only the GB atoms in the pure Cu polycrystal (i.e., all crystalline atoms have been removed), with each atom colored according to the disorientation angle of the GB to which it belongs. Figure 2(b) presents the Cu-Zr polycrystal with a similar coloring scheme for the GB atoms, except with Zr dopant atoms being colored black. The distribution of disorientation angles between the neighboring grains for both polycrystalline Cu (black circles) and Cu-Zr (blue squares) is shown in Figure 2(c). The GB disorientation angle in both of the polycrystalline samples ranges from 10° to 60°, covering nearly the entire angle range expected for a randomly textured polycrystalline sample. The disorientation angle distributions in both samples first increase then decrease, following the Mackenzie distribution which is expected for a random polycrystal [81]. To determine if the GB networks are similar between the pure and doped samples, we compare the disorientation angles associated with specific boundaries in Figure 2(d). The data points reside close to the green reflection line which indicates that GB disorientations have not changed significantly due to the segregation process. Thus, with both the mean grain size and the distribution of GB disorientation matching in the pure and alloyed samples at 900 K, the GB descriptors from the pure Cu sample can be used as the reference point for understanding how segregation is influenced by the starting GB structure.

One method of measuring GB segregation is to probe the dopant concentration profile across the boundary thickness. For example, Hu et al. measured the concentration profile across an amorphous complexion in a Cu-Zr thin film with energy-dispersive X-ray spectroscopy inside of a transmission electron microscope, finding a maximum Zr content of 6 at.% at the GB [82]. Similarly, Zheng et al. [83] used atom probe tomography to measure concentration profile across the GB thickness in a Ni-Bi alloy, reporting a maximum GB solute content of 51.4 at.%. Another approach to determine interfacial segregation is to measure solute excess of the GBs [84-86], which can be compared to the excess concentration of solute as expressed by the Gibbs adsorption isotherm and defined as the difference between the amount of solute present per unit area of GB and that which is present in the bulk. For example, a recent study by Gupta et al. [87] used GB excess to quantify P segregation in a Ni-P alloy using both atom probe tomography and atomistic simulations. Here, we use both metrics to assess the segregation tendency of different GBs, with atomic visualizations of each presented in Figure 3. Figure 3(a) shows the GB atoms in the Cu-Zr polycrystal with atoms colored according to the average concentration of Zr atoms within that given GB section. Although, the global Zr concentration in Cu polycrystal is 2 at.%, locally the Zr content at the GBs was observed to vary from very low, near 0 at.% Zr (black), to very high concentrations, near 20 at.% Zr (white). Figure 3(b) also shows only the GB atoms in the same Cu-Zr polycrystal, although this time the atoms are colored according to the GB excess of a given GB. The interfacial excess varies from 0 to 14 atoms/nm$^2$. For the most part, GBs with high GB excess also have high solute concentration, although a few exceptions are observed. One such exception is marked with a black arrow. This GB has a moderate GB concentration (10.8 at.% Zr, or roughly in the middle of the observed range) but a high GB excess (12.7 atoms/nm$^2$, or close to the maximum of the observed range). Careful inspection shows that this particular boundary has a thickness of ~1.3 nm, which is much thicker than most of the boundaries in the alloy sample (GB thickness is explored in more detail in Section 3.2). The thicker boundary has more Zr dopants overall, leading to a large GB excess, yet the mean composition of this boundary is in fact modest. This example shows that although GB excess can be an important parameter to study GB segregation, it cannot account for significant variations in GB thickness.

Figure 4(a) quantifies the distribution of GB solute concentration in the Cu-Zr alloy at 900 K in histogram form. Figure 4(b) presents a cumulative distribution function of the same data, which quantifies the fraction of the total number of GBs that have a GB concentration lower than a given value. For example, this curve shows that 90% of the GBs in the Cu-Zr alloy at 900 K have GB concentrations lower than ~15 at.% Zr. The range of GB composition varies from 0.7 at.% to 19.5 at.% Zr, with a peak in the histogram data located at a solute concentration near ~13 at.% Zr. The histogram in Figure 4(a) reveals that the distribution of GB concentration is not symmetric, with an extended tail into lower composition values (the extended tail at low values in Figure 4(b) shows this same feature). In fact, a normal or Gaussian fit is unable to capture the shape of the GB concentration distribution, necessitating a more complex fitting procedure. This is in contrast to the machine learning approach used by Huber et al. [88] to model GB segregation in bicrystal samples, where the segregation energy density of states was forced to fit a Gaussian distribution. To describe the shape of the GB concentration distribution, we instead use a skew-normal model to capture the different tails of the data. A skew-normal distribution can be thought of as a Gaussian model that is modified to allow the shape to become asymmetric, and is described by three parameters: (1) the location, $\mu$, (2) the scale, $\sigma$, and (3) the shape, $\alpha$. For perspective, well known terms mean and standard deviation refer to the location and scale of a Gaussian fit. The shape parameter describes the level of skewness needed to fit the data, with a negative value signaling a distribution tail that extends to lower values, as shown in Figure 4(a) where $\alpha = -4.5$ for the GB concentration data. This skew-normal fit works well to describe the data in this study, with high coefficient of determination, $R^2$, values found for all measures of GB segregation. Interestingly, this success of a skew-normal model for describing the GB concentration distribution within a polycrystalline ensemble mimics the recent success of Wagih and Schuh [41, 42] in using such a model to describe the full atomic spectrum of segregation energies in an Al-Mg alloy. These authors investigated the segregation energy for Mg dopants at every atomic GB site, finding possibilities ranging from segregation (negative energy) to anti-segregation (positive energy), with the overall distribution fit well by a skew-normal model shifted toward segregation behavior. Wagih and Schuh noted that the most favorable atomic segregation sites were found to be well-distributed throughout the GB network in their simulations, with no obvious clustering at specific boundaries. The persistence of the skew-normal shape to the GB concentration observations of this study, where we analyze a mesoscale behavior as each GB concentration value is the mean from a given GB section which itself contains many possible atomic segregation sites, demonstrates that certain boundaries are doped at a faster rate than others.

To provide a second metric of GB segregation, Figures 4(c) and 4(d) show the distribution of GB excess in the Cu-Zr alloy at 900 K in a histogram and cumulative distribution function format, respectively. The peak of the GB excess histogram occurs near ~6 atoms/nm$^2$ and the range of the distribution goes fully from 0 to 14 atoms/nm$^2$. Similar to the GB concentration data, the GB excess can be well fit with a skew-normal model, again with a negative shape ($\alpha = -4.0$) to account for a tail toward lower GB excess values. We note that although the skew-normal model generally works here, it appears to systematically miss the extremely high values of GB excess. As mentioned previously, these high values appear to be associated with a small population of GBs that have grown thicker than usual. For the remainder of this paper, we focus on GB concentration as the two metrics largely show the same trends. As a whole, the observations of Figure 4 demonstrate that all GBs are not the same, as a wide variety of GB concentrations are observed, and GBs can therefore not be thought of as a single type of defect when considering segregation phenomena. The existing McLean-type segregation models are focused on single GBs with common structure as compared to a general collection of different GBs observed in a polycrystal. For example, these types of simplified models may predict solute segregation within dilute limits at symmetric tilt boundaries based on a GB parameter such as misorientation angle or free volume. However, such models are misleading for a polycrystalline alloy where segregation is inhomogeneous across the interfacial network, with different boundaries having different solute concentrations. Even a normal fitting model with mean and standard deviation is not enough to describe GB concentration, as important tails in the distribution require a skewness for a good fit.

Armed with a description of the interfacial concentrations (Figure 4) and noting that there has been very little to no overall change of the grain shapes, sizes, and orientations for the Cu-Zr alloy at 900 K as compared to the pure Cu system (Figure 2), we next investigate whether the observed segregation trends can be understood in terms of the starting local boundary structure or its evolution with doping. Crystallography of a GB can potentially influence segregation [37, 89], with a commonly used metric being the disorientation between grains across a specific GB. Segregation fundamentally occurs because the dopant atoms prefer to reside in the GB region, with possible causes being the desire to fill a high energy site (see, e.g., [90] ) or an atomic site with excess/free volume (see, e.g., [91]). GB energy can be defined as the difference between the potential energy of GB atoms and the potential energy of grain interior atoms, and in this case is expressed as a per atom quantity here. GB free volume can be defined as the difference between the Voronoi volume of GB atoms and the same volume of grain interior atoms. Figure 5 presents the measured GB concentrations of Zr as functions of the initial boundary character (disorientation), energy (GB energy), and structure (GB free volume). In Figure 5(a), the GB concentration shows no clear trend with disorientation. Figures 5(b) and 5(c) show that GB concentration generally increases with increasing GB energy and GB free volume, respectively, but these correlations are relatively weak. The weak correlation is evidenced by the low $R^2$ values for the curve fits, but can also be visualized by noticing the wide range of concentration that can be found for a given value for disorientation, GB energy, or GB free volume, as denoted by the red arrows in each figure part. For example, for a GB free volume of 0.6 $Å^3$/atom, the observed GB concentration can vary from 5-18 at.% Zr in Figure 5(c), spanning a large percentage of the overall observed range of concentration. As such, it is clear that the observed trends in segregation cannot be adequately predicted by starting boundary state.

We next move to understand if the evolution of boundary state is reflective of the segregation that is occurring. To capture this, the changes to GB energy and GB free volume are probed, calculated for a given boundary as the value in the alloy system minus the value in the pure Cu reference system, and presented in Figure 6. GB energy generally decreases with increasing Zr concentration, demonstrating that GB doping reduces the energetic penalty associated with the presence of a GB defect. In contrast, the GB free volume generally increases with increasing Zr concentration at the boundary, which can be connected to the fact that Zr is a larger atom than Cu and segregation leads to larger Voronoi volume associated with a given GB site. While the fit of the regression curves shown in Figure 6 is better than that in Figure 5, noticeable boundary-to-boundary variations are still observed in the measure GB structure or energy for a given GB concentration value. As a whole, this discussion emphasizes that boundary-to-boundary variations in GB segregation phenomena are large in a random polycrystal, and that local nuances of GB state should be important for determining both what level of segregation occurs and how that segregation alters the GB.

Next, we examine GB segregation in the Cu-Zr alloy at 1050 K, to investigate the effect of increasing temperature into a range where many GBs transition to amorphous complexions (to be shown explicitly in Section 3.2). Cross-sectional snapshots colored by local chemical composition are presented in Figures 7(a) and 7(b) to show segregation at a representative collection of GBs in Cu-Zr polycrystals at 900 K and 1050 K, with the images taken along a (111) viewing plane according to the sample coordinate system shown in Figure 1. The local/atomic Zr concentration is determined by taking a sphere of radius 0.3 nm around the reference atom, large enough to include at least one neighbor shell for each chemical species in the system [92]. The local concentration of Zr is then determined as a ratio of the number of Zr atoms and the total number of atoms in the sphere. Solute atoms are primarily segregated in the GB network (denoted by black dashed lines) for both samples, although a few GBs experience very little Zr segregation. Dopant concentration in the grains is very low, as shown by the blue colored atoms in the grain interiors. At 900 K (Figure 7(a)), the majority of the GB network is moderately doped (green) along with only a few solute rich regions (red). Whereas at 1050 K (Figure 7(b)), a higher number of solute rich regions are found within the GB network. We do note that the average grain size of the Cu-Zr polycrystal has increased slightly to 10.9 nm upon increasing the temperature to 1050 K, and noticeable grain structure rearrangement has occurred, precluding a direct comparison on individual boundaries. However, an overall comparison between the alloy at the two temperatures is possible by again quantifying the average concentration of Zr for each distinct GB and investigating the resulting distribution for the polycrystalline aggregate. Figures 7(c) and 7(d) present the distribution of GB dopant concentration for the Cu-Zr alloy at 1050 K in histogram and cumulative distribution function format, respectively. Similar to the 900 K sample, the GB concentration at a given boundary can vary over a wide range, with the lower end of the population being below the global composition. However, the maximum observed GB concentration is actually lower for the 1050 K sample (~16 at.% Zr) than it was for the 900 K sample (~20 at.% Zr). A possible reason for decreasing GB concentration with increasing temperature could be an increase in the solubility of the grains. However, we find that grain interior concentration at 1050 K is ~0.2 at.% Zr, as compared to ~0.1 at.% Zr at 900 K. This means that the changing GB concentration is not the result of desegregation, but rather redistribution of the dopant atoms within the GB network. The GB concentration distribution is more closely clustered around the more heavily doped values, as evidenced by the scale of the distribution for 1050 K ($\sigma = 2.9$) being smaller than that for 900 K ($\sigma = 5.0$), which corresponds to less spread of the data. A side effect of this reduced spread is that the distribution requires a larger shape/skew value ($\alpha = -6.0$) to capture the GBs with less segregating solute. Such a decrease in GB concentration variation is also observed by Lu et al. [44], where a Pt-Au thin film equilibrated at 773 K had a large variation in Au concentration at the GBs as compared to a Pt-Au thin film equilibrated at 973 K, which had a uniform solute concentration at the GBs. We hypothesize that the shift to a less varied concentration distribution at higher temperature is inherently related to complexion transitions, which will be explored in the next section.

**3.2 GB complexion coexistence**

Heavily doped boundaries have been observed to undergo premelting to form amorphous complexions as temperature is increased and a local liquid-like structure is formed along the interface (see, e.g., [93-95]), so we next work to identify and quantify the complexion distributions in the polycrystals studied here. While there are multiple ways that complexions can be categorized, thickness is a metric which is commonly used to describe amorphous complexions and is experimentally viable, so it is chosen here as our focus. Figure 8(a) presents GB thickness measurements from the Cu-Zr polycrystals at 900 K and 1050 K, with the range of thicknesses of the pure Cu at 900 K shown as a grey region. The pure Cu and Cu-Zr polycrystals contain nearly identical numbers of GBs at 900 K (~470 GBs), whereas the number of GBs has decreased by ~20% at 1050 K (~370 GBs) due to slight coarsening of the grain structure. GB thickness in the pure Cu polycrystal varies from 0.45 nm to 0.85 nm, so this range can serve as a reference state that describes the expected range of thickness for only ordered GBs. We note that individual GB atoms will still be identified as defective even in very thin and well-defined boundaries by atomistic analysis techniques such as CNA or the bond-orientational order parameter [96]. GBs with thicknesses outside of the range observed in the pure Cu sample are candidates for complexion transitions and merit further inspection. GB thickness in the Cu-Zr alloy at 900 K varies from 0.2 nm to 1.7 nm, with the data represented by blue circles in Figure 8(a). Only a small fraction of the GBs have thicknesses greater than the 0.85 nm maximum GB thickness observed in pure Cu at the same temperature, indicating minor evolution due to segregation at this temperature. In contrast, significantly more evolution of the GB thickness distribution is observed for the Cu-Zr alloy at 1050 K, where GB thickness (represented by black squares in Figure 8(a)) varies from 0.2 nm to 3.6 nm. We note that some of the GBs in this sample were not automatically identified by the GTA code due to a drastic increase in their thickness, as the code had difficulty assigning the neighboring grains for thick complexions. These GBs were identified manually by analyzing their atomic structure using CNA in OVITO and then added to the dataset. While not discussed previously, this procedure was in fact performed before extracting the GB composition data shown in the last section, so all figures in this work represent the complete GB network.

Figure 8(b) shows a representative example of a doped yet still structurally ordered GB that is 0.7 nm thick in Cu-Zr at 1050 K. In contrast, Figure 8(c) shows a representative example of an amorphous complexion with 2 nm thickness GB complexion in the same polycrystal. These two example boundaries are denoted in Figure 8(a) as well, supporting the concept that GBs with thicknesses in the reference range can be considered *ordered GBs*, while those with larger thickness are *amorphous complexions*. Inspection of Figure 8(a) clearly shows that a larger number of GBs have transformed to amorphous complexions in the alloy sample at 1050 K. Figure 8(a) also shows that there is a small but noticeable population of GBs which are thinner than the range of reference data, with 4 and 21 thinner boundaries identified for the alloy at 900 K and 1050 K, respectively. Figure 9(a) presents a view of these thin boundaries in the sample at 1050 K, where they appear as thin planes and are colored according to their GB number from the GTA software analysis. Further inspection demonstrates that these features are not actually traditional GBs but instead annealing twins that form inside of individual grains. Figure 9(b) shows a representative example, where FCC atoms appear green, HCP atoms appear red, and other/GB atoms appear white. Annealing twins are common in Cu systems due to the low stacking fault energy and these interfaces have negligible segregation (less than 0.1 at.% Zr), which is comparable to the concentration measured in the grain interior regions. As such, we do not include these boundaries in subsequent analysis of GB thickness, and these features were in fact removed when presenting earlier data on GB concentration.

Figure 10(a) presents the GB thickness data for all three samples in cumulative distribution form. The cumulative fraction curve of the pure Cu polycrystal at 900 K (black triangles) is on the far left, and can again be taken as a reference state for a collection of ordered GBs and the expected variations in thickness for a "normal" GB network. The Cu-Zr alloy at 900 K (blue circles) largely mimics the GB thickness distribution of the pure Cu and has only a small subset of GBs (~6%) which are above the maximum thickness of the pure system. On the other hand, the Cu-Zr alloy at 1050 K (red squares) has a significant subset of boundaries that are further to the right in Figure 10(a), demonstrating that ~23% of the boundaries in the GB network are amorphous complexions. These observations are consistent with experimental reports of amorphous complexions in the literature, as GB segregation is needed for boundary premelting below the bulk solidus temperature and higher temperatures promote such a GB complexion transition [57]. However, while experimental reports on amorphous complexion thicknesses are reported in Refs. [47, 79], these studies were not able to simultaneously quantify the full distribution of ordered interfaces in the same GB network due to experimental limitations, meaning the measurement of the transformed fraction, as done here, was not possible. The thicker amorphous films are easier to isolate and then measure, while it is nearly impossible to identify every single ordered boundary in the system and also more difficult to measure thickness for those interfaces when examples are identified. The boundaries which have undergone a complexion transition to an amorphous structure (i.e., the GBs thicker than the pure Cu reference state) were isolated by truncating the GB thickness data at the black dash line in Figure 10(a), with the thickness distributions for the amorphous complexions only shown in Figure 10(b). These updated distributions are compared to experimental amorphous complexion thickness distributions in Cu-Zr alloys from Grigorian and Rupert [79], shown in grey, with the different curves corresponding to different quenching rates from 1223 K (950 °C). For this experimental data, the fastest quenching rate (curve furthest to the right) corresponds to the amorphous complexion distribution representative of the true high temperature state, while the other curves represent GB networks where some of the amorphous complexions have reverted back to ordered GBs during cooling [79]. Comparison of the experimental data against computationally-obtained distributions containing more information, such as the results from this study, can allow for features such as the transformed fraction to be quantified in the future. For example, Figure 10(b) suggests that even at the slowest quenching rate shown here, more than 6% of the total boundaries are amorphous complexions because the experimental curve lies to the right of the blue simulated curve. Similarly, the three curves for faster quench rates all lie to the right of the red simulated curve, suggesting that they all have more than 23% of the total boundaries with an amorphous complexion structure. Accurate extraction of the exact transformed fraction for each experimental curve would require many more polycrystalline GB network simulations to capture the full spectrum of possibilities, but this discussion demonstrates the potential utility of having complete computational datasets describing GB network structure.

The separate populations of GB thickness shown in Figure 10 demonstrate that two types of complexions can coexist within a polycrystalline GB network, supporting earlier experimental reports where some boundaries were thin and ordered while others formed amorphous films with uniform thickness [19]. However, in an experimental study, one cannot ensure that the system is in equilibrium and also the cooling rate from the elevated temperature is important, as shown by Grigorian and Rupert [79]. This work clearly shows that a polycrystalline sample can find an equilibrium state where both ordered and amorphous complexions coexist. With this complexion coexistence in mind, we revisit the GB concentration data shown in Section 3.1, but this time separate the information into sub-populations, with one corresponding to only the ordered GBs and another to the amorphous complexions. Figure 11 presents this separated GB concentration data for the simulated Cu-Zr alloy at the two different temperatures. Similar to the complete dataset, the individual complexion datasets can be well fit with skew-normal models in each case, providing further evidence of two distinct complexion sub-populations that coexist within a single GB network. We note that the GB concentration of the ordered GBs in the sample at 1050 K (Figure 11(b)) has the highest skew to the model fit ($\alpha$ = -8.6), with this skewed shape propagating to cause the large skew previously observed in Figure 7(c) for the GB composition of all boundaries. As such, the overall shape of the GB concentration data for the higher temperature is actually dominated by the ordered GB sub-population. For both temperatures, the amorphous complexions have lower shape values, indicating less spread around the most common value or a tighter distribution. In other words, the amorphous complexions tend to have Zr concentrations near a common mean value, while the ordered complexions demonstrate much more variation. While heavily-doped ordered complexions exist in the system, ordered complexions with almost no dopant atoms are found as well. In addition, we observe that the GBs with the highest concentration in the 900 K sample are notably ordered GBs, demonstrating that the most heavily doped boundaries are not necessarily those which transform to amorphous complexions. More research is needed to understand such complexion transitions and their connection to GB structure in order to fully understand these variations, but some commonly held assumptions can be called into question. For example, early thermodynamic models assumed a common boundary concentration at which complexion transitions occurred [93, 97, 98]. Such a description might be acceptable for limited subsets of boundaries, for example, where a common critical grain boundary composition for amorphization was observed for a set of symmetric tilt and twist boundaries [65]. However, random polycrystals appear to exhibit more complexity and will necessitate more nuanced theories to connect grain boundary chemistry to complexion structure for different boundary character.

For the data described in Figure 11, a GB concentration is assigned to each identified GB, yet it is also possible that there are local variations in composition within a given boundary. Figure 12(a) shows the Cu-Zr polycrystal at 1050 K, with examples of ordered GBs marked by black circles and examples of amorphous complexions denoted by red circles. Zoomed views of each of these boundaries are shown in Figures 12(b) and (c), where atoms are colored according to their local/atomic Zr concentration. For all of the GBs, the local Zr concentration is significantly higher than the grain interiors, with the most noticeable difference between the two complexion types being the presence of multiple solute rich clusters in the amorphous complexions. The gradients in color within the amorphous complexions in Figure 12(c) correspond to composition heterogeneities within a given boundary. Figure 13(a) shows atomic-scale measurements of Zr concentration for five examples each of ordered GBs and amorphous complexions, where the mean values (data reported in Figure 11) for each boundary are represented by the large black circle and the individual local concentration values within that boundary are represented by colored X (ordered GBs) or + (amorphous complexions) symbols. We note that these are only examples to demonstrate the idea to the reader, and Figure 11 should be used to draw conclusions about the relative GB concentrations of the different complexion types. In all cases, significant local variations are observed, which we quantify with the local *concentration range* (maximum minus minimum local/atomic concentration within a given GB). Figure 13(b) shows the distributions and average values of the concentration range for the two complexion types in the samples at the two different temperatures, where the mean values for each complexion type are represented by the large black circles and the individual concentration range values for that complexion are represented by blue (ordered GBs) and red (amorphous complexions) colored data points. Each of the data points is therefore the range of compositions within a given GB. At 900 K, the average concentration range was found to be larger at amorphous complexions (12.7 at.% Zr) than at ordered GBs (10.4 at.% Zr). Similar behavior is observed in the higher temperature sample where the average concentration ranges for amorphous complexions and ordered GBs are 15.1 at.% Zr and 12.3 at.% Zr, respectively. While Figure 11 (where each data point represents the average concentration of a given boundary) showed that the distribution of average composition is tighter for the amorphous complexions, Figure 13 (where each data point represents atomic-level chemical composition) shows that the amorphous complexions generally have more heterogeneity on a local length scale. As a whole, the comparison between these two figures highlights the important differences which can be observed within individual boundaries, as well as within populations of a given complexion type.

In general, our observations of complexion co-existence and various segregation trends emphasize the intricacies associated with GB segregation and complexion transitions within a polycrystalline network. A major takeaway is that the community must begin to incorporate this complexity into models of such behavior. For example, Zhou and Luo [99] created equilibrium complexion diagrams for multi-component alloys to describe the thermodynamic conditions where complexion transitions occurred. While this work represented a major advancement on this topic, the complexion diagrams only incorporated $\pm 15\%$ variations in GB energy to account for anisotropy in a polycrystal [100], while our results show much greater variety in local composition within a given boundary, mean GB concentration distributions, and complexion thicknesses within a polycrystalline GB network. It is important to note that a random polycrystal is studied here, meaning that the variety of GB descriptors needed is representative of the diversity which can be expected for a set of general GBs. In addition, the simulated polycrystals studied here allow for the investigation of a large number of GBs as compared to experimental techniques such as transmission electron microscopy or atom probe tomography. We finish by noting that the integration of such experimental techniques with computational studies will likely provide a more comprehensive description of these important microstructures, with the fraction of boundaries transformed to a given complexion state being an obvious target for such an integrated study.

## 4. Summary and conclusions

In this work, atomic details of solute segregation and segregation induced structural transitions within a polycrystalline GB network are studied using hybrid atomistic simulations. A polycrystalline pure Cu and Cu-Zr alloy are used as model systems to study the correlations between different GB parameters and the transition of ordered interfaces to amorphous complexions. The following conclusions can be drawn from this study:

- Minimal changes in the grain structure are observed upon solute segregation in pure Cu polycrystalline sample at 900 K. Boundaries with high GB excess generally correlate with high GB concentration, although a few exceptions exist where moderately doped GBs have high GB excess. These interfaces are found to be thicker than average, indicating that significant variations in GB thickness can convolute GB excess as a useful metric.

- Large variations in interfacial segregation are observed with skew-normal models capturing the asymmetry in segregation behavior across the boundary network. The GB concentration distribution at 1050 K is heavily clustered around the high GB concentration values with a larger skew in the distribution as compared to the 900 K sample, which is related to a greater number of complexion transitions at the higher temperature.

- GBs with thickness greater than 0.85 nm, based on the upper limit of GB thickness in pure Cu, are identified as amorphous complexions. At 900 K, only 6% of GBs in Cu-Zr alloy are amorphous complexions indicating very few transitions upon segregation at this temperature. In contrast, significant changes in the grain structure are observed at 1050 K, where 23% of GBs are found to be amorphous complexions, demonstrating the tendency for increased frequency of complexion transitions at higher temperatures. Computationally obtained distributions of transformed boundaries in polycrystal models provide useful benchmarks to compare against experimental data where it is not possible to isolate all the GBs due to practical limitations.
- The coexistence of two different complexion populations within a single GB network can be observed (ordered versus amorphous), with each having its own distinct skew-normal distribution of GB concentration. GBs with the highest concentration in the 900 K sample are notably ordered GBs, demonstrating that the most heavily doped boundaries are not necessarily those which transform to amorphous complexions.

Overall, this work provides a more complete understanding of solute segregation behavior and subsequent structural transitions in a polycrystalline GB network. The high resolution offered by atomistic simulations makes it possible to characterize segregation at GB network in a polycrystalline system, which is not available in bicrystal geometries. This complex set of information improves the field's understanding of other general defect structures in the GB network and highlights that microstructural complexity must be treated in detail to allow for tunable alloy microstructure and properties.

**Acknowledgments**

This research was supported by the U.S. Department of Energy, Office of Science, Basic Energy Sciences, under Award No. DE-SC0021224. Structural analysis and atomic visualization were performed with software funded by the National Science Foundation Materials Research Science and Engineering Center program through the UC Irvine Center for Complex and Active Materials (DMR-2011967). Z.P. acknowledges help from Dr. Jason Panzarino with running the Grain Tracking Algorithm (GTA) code.

# References

[1] M.A. Meyers, A. Mishra, D.J. Benson, Mechanical properties of nanocrystalline materials, Progress in Materials Science 51(4) (2006) 427-556.

[2] K. Kumar, H. Van Swygenhoven, S. Suresh, Mechanical behavior of nanocrystalline metals and alloys, Acta Materialia 51(19) (2003) 5743-5774.

[3] M. Dao, L. Lu, R. Asaro, J.T.M. De Hosson, E. Ma, Toward a quantitative understanding of mechanical behavior of nanocrystalline metals, Acta Materialia 55(12) (2007) 4041-4065.

[4] J.A. Sharon, H.A. Padilla, B.L. Boyce, Interpreting the ductility of nanocrystalline metals 1, Journal of Materials Research 28(12) (2013) 1539.

[5] S.N. Mathaudhu, B.L. Boyce, Thermal stability: the next frontier for nanocrystalline materials, JOM 67(12) (2015) 2785-2787.

[6] A.R. Kalidindi, T. Chookajorn, C.A. Schuh, Nanocrystalline materials at equilibrium: a thermodynamic review, JOM 67(12) (2015) 2834-2843.

[7] H. Peng, M. Gong, Y. Chen, F. Liu, Thermal stability of nanocrystalline materials: thermodynamics and kinetics, International Materials Reviews 62(6) (2017) 303-333.

[8] Y. Hu, T.J. Rupert, Atomistic modeling of interfacial segregation and structural transitions in ternary alloys, Journal of Materials Science 54(5) (2019) 3975-3993.

[9] S. Turnage, M. Rajagopalan, K. Darling, P. Garg, C. Kale, B. Bazehhour, I. Adlakha, B. Hornbuckle, C. Williams, P. Peralta, Anomalous mechanical behavior of nanocrystalline binary alloys under extreme conditions, Nature Communications 9(1) (2018) 1-10.

[10] T.J. Rupert, Solid solution strengthening and softening due to collective nanocrystalline deformation physics, Scripta Materialia 81 (2014) 44-47.

[11] B. Hornbuckle, T. Rojhirunsakool, M. Rajagopalan, T. Alam, G.P. Pun, R. Banerjee, K. Solanki, Y. Mishin, L. Kecskes, K. Darling, Effect of Ta solute concentration on the microstructural evolution in immiscible Cu-Ta alloys, JOM 67(12) (2015) 2802-2809.

[12] B.C. Hornbuckle, C.L. Williams, S.W. Dean, X. Zhou, C. Kale, S.A. Turnage, J.D. Clayton, G.B. Thompson, A.K. Giri, K.N. Solanki, Stable microstructure in a nanocrystalline copper–tantalum alloy during shock loading, Communications Materials 1(1) (2020) 1-6.

[13] K. Darling, A. Roberts, Y. Mishin, S. Mathaudhu, L. Kecskes, Grain size stabilization of nanocrystalline copper at high temperatures by alloying with tantalum, Journal of Alloys and Compounds 573 (2013) 142-150.

[14] M. Rajagopalan, K. Darling, C. Kale, S. Turnage, R. Koju, B. Hornbuckle, Y. Mishin, K. Solanki, Nanotechnology enabled design of a structural material with extreme strength as well as thermal and electrical properties, Materials Today 31 (2019) 10-20.

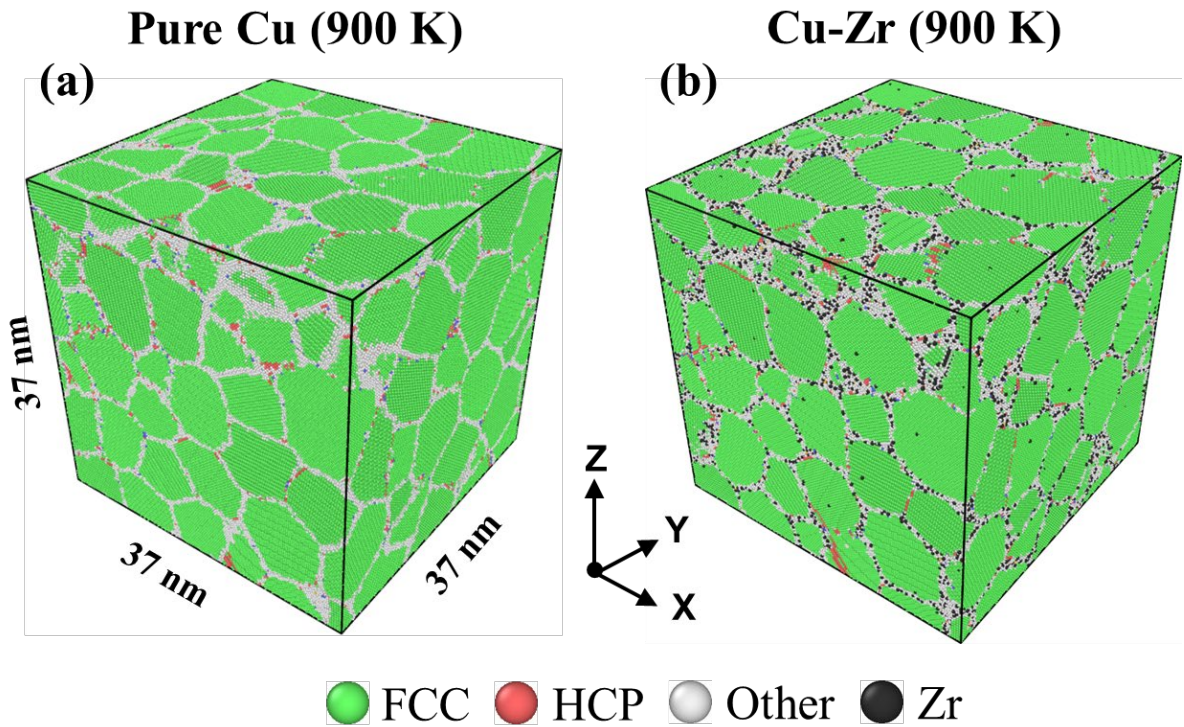

**Figure 1.** Polycrystalline samples of (a) pure Cu equilibrated at 900 K with an average grain size of 10.1 nm and (b) Cu-2 at.% Zr equilibrated at 900 K with an average grain size of 10.2 nm. Cu atoms are colored according to their local atomic structure (FCC, HCP, or other), while Zr atoms are colored black.

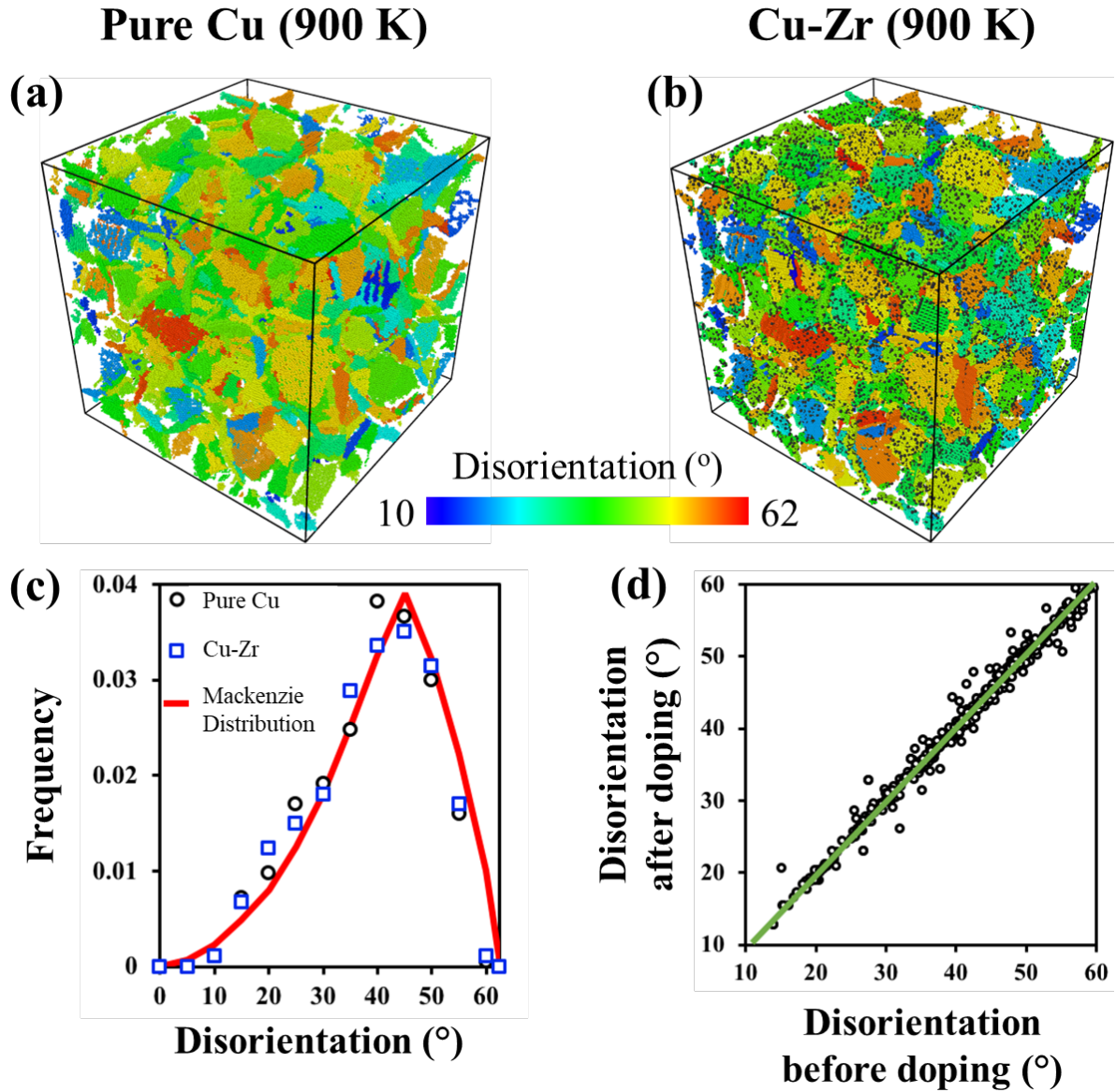

**Figure 2.** GB networks in polycrystalline samples of (a) pure Cu equilibrated at 900 K and (b) Cu-2 at.% Zr equilibrated at 900 K, with boundary atoms colored according to the disorientation angle across the interface. (c) The distributions of GB disorientation angle before and after doping, as well as the Mackenzie distribution for reference, demonstrating that a random polycrystalline grain structure is found in both cases. (d) The disorientation angle of each boundary after doping plotted against the disorientation angle of that same boundary before doping. Data points reside near the green 1-to-1 line, indicating little to no structural evolution during the segregation process.

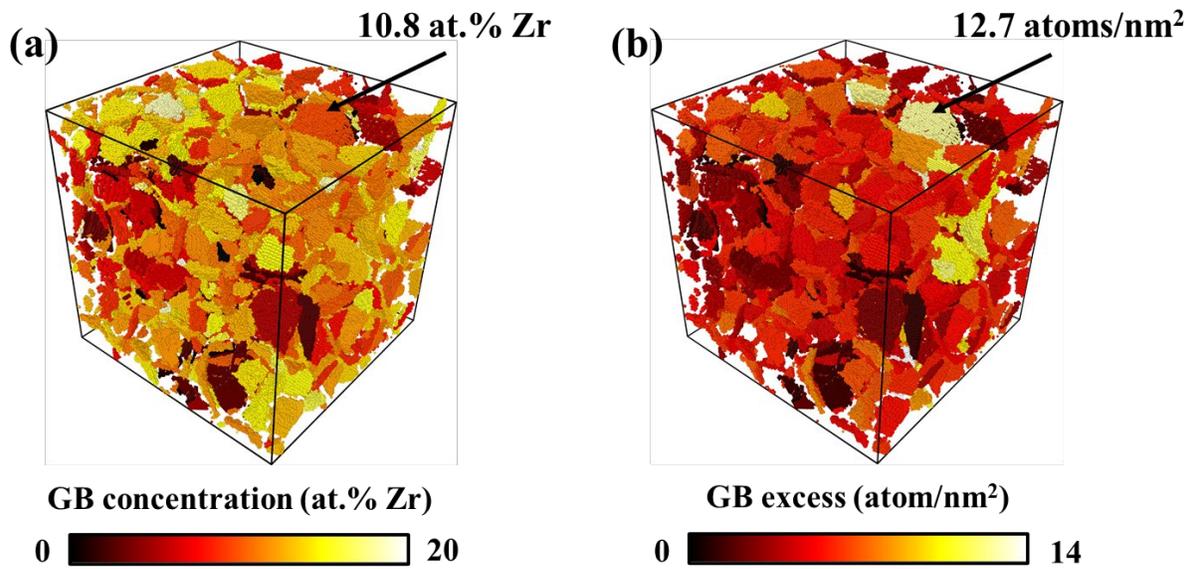

**Figure 3.** Segregation tendency of different GBs in the Cu-Zr polycrystal equilibrated at 900 K. Only the boundary atoms are shown and are colored according to (a) the local GB concentration of Zr and (b) the GB excess within each interface. An example of a GB with moderate dopant concentration but high grain boundary excess is denoted by the black arrow. This interface is thicker than the average boundary, demonstrating how interface thickness can cause the two segregation metrics to differ.

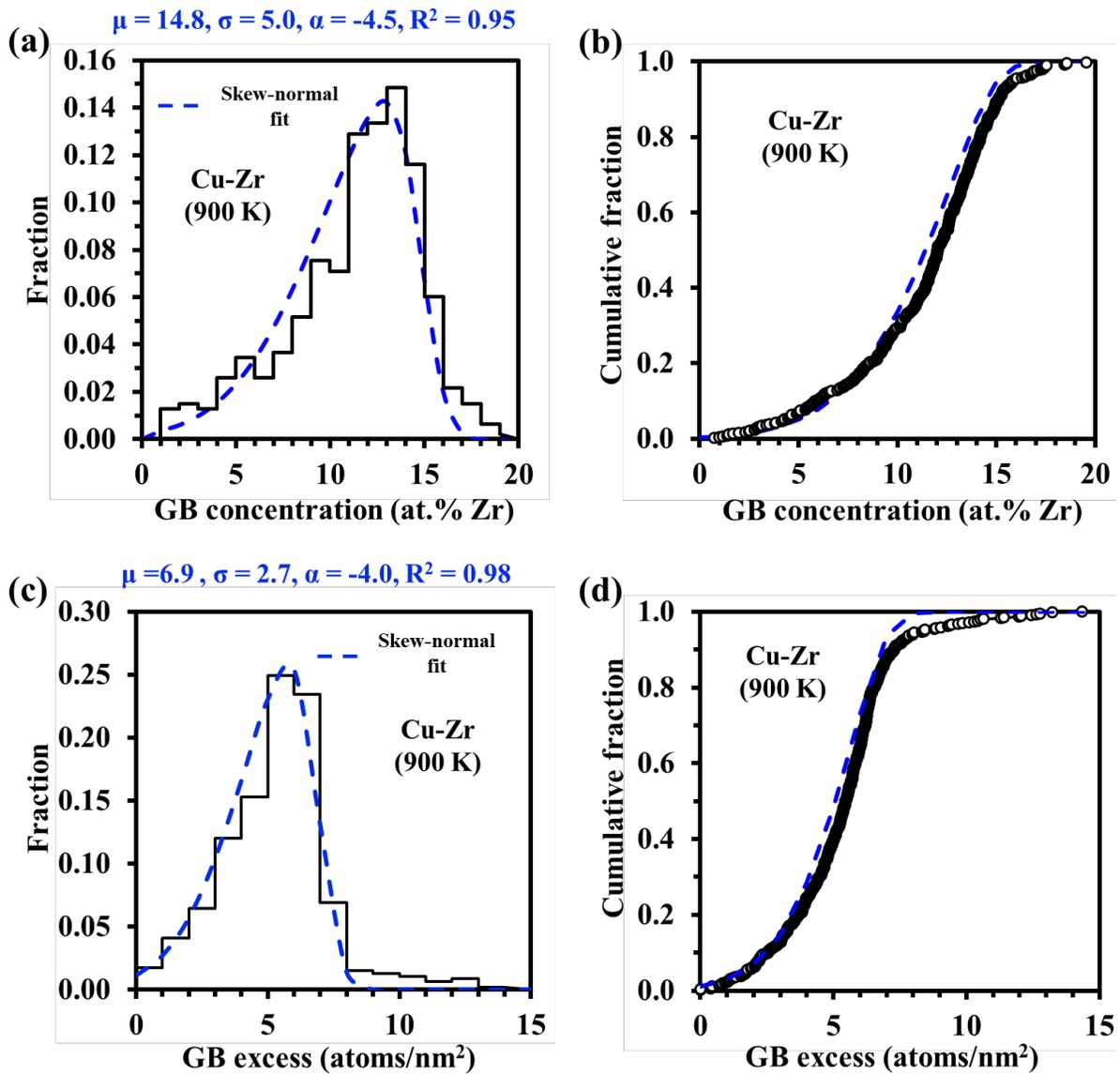

**Figure 4.** The statistical populations of (a)-(b) GB solute concentration and (c)-(d) GB excess in the Cu-Zr polycrystal at 900 K. In (a) and (c), data from this study is shown in histogram form while a skew-normal fit is also presented, showing reasonable agreement with the shape of the data. A cumulative distribution function of this same data is shown in (b) and (d).

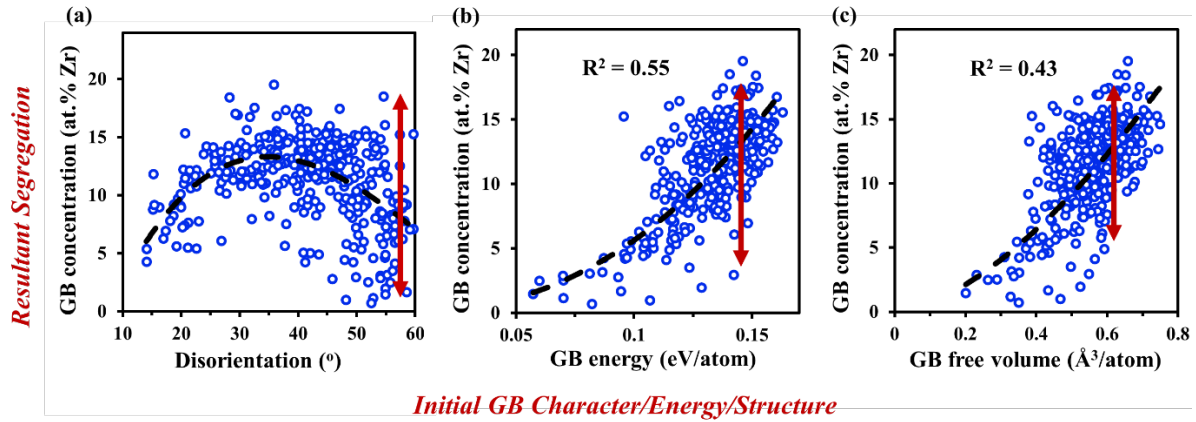

**Figure 5.** GB concentration plotted against the initial character, energy, and structure of each boundary, using (a) disorientation angle, (b) GB energy, and (c) GB free volume as metrics for these features, respectively. No predictive correlations are observed, with large variations in GB concentration found for a single chosen value of any of these parameters.

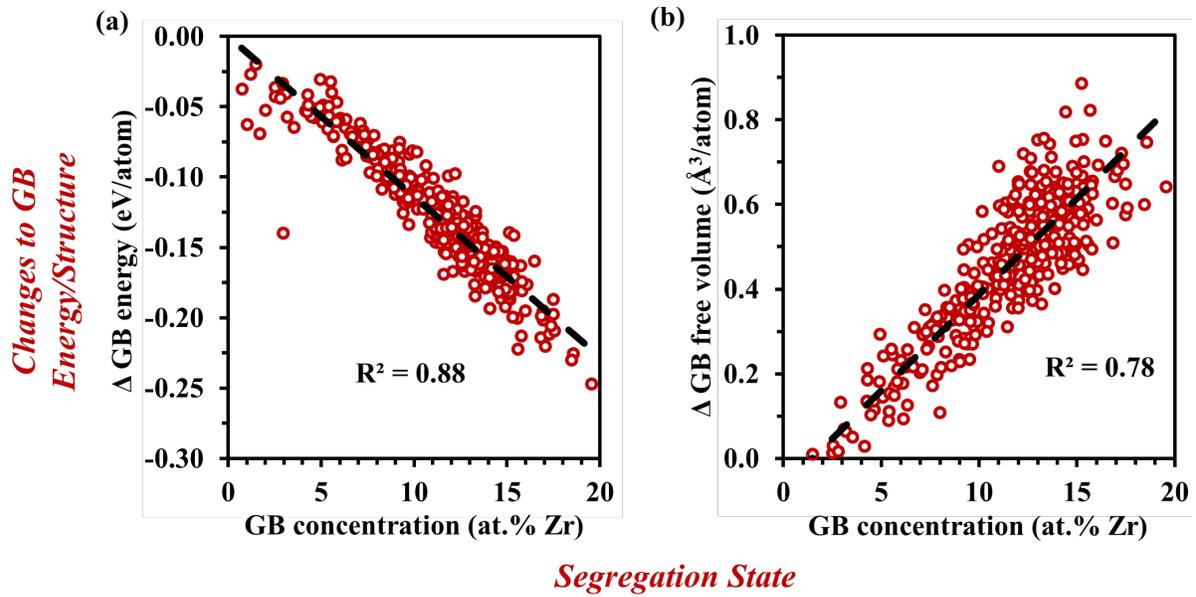

**Figure 6.** The effect of doping on GB energy and structure, as shown by correlations between the observed segregation state and physical changes to the GBs in the Cu-Zr polycrystal equilibrated at 900 K. (a) Change in GB energy and (b) change in GB free volume, both measured for a given interface in the alloy using the same interface in the pure Cu sample as a reference point, are plotted as a function of GB concentration. The change in free volume monotonically increases with increasing GB concentration, while the change in boundary energy monotonically decreases with increasing GB concentration.

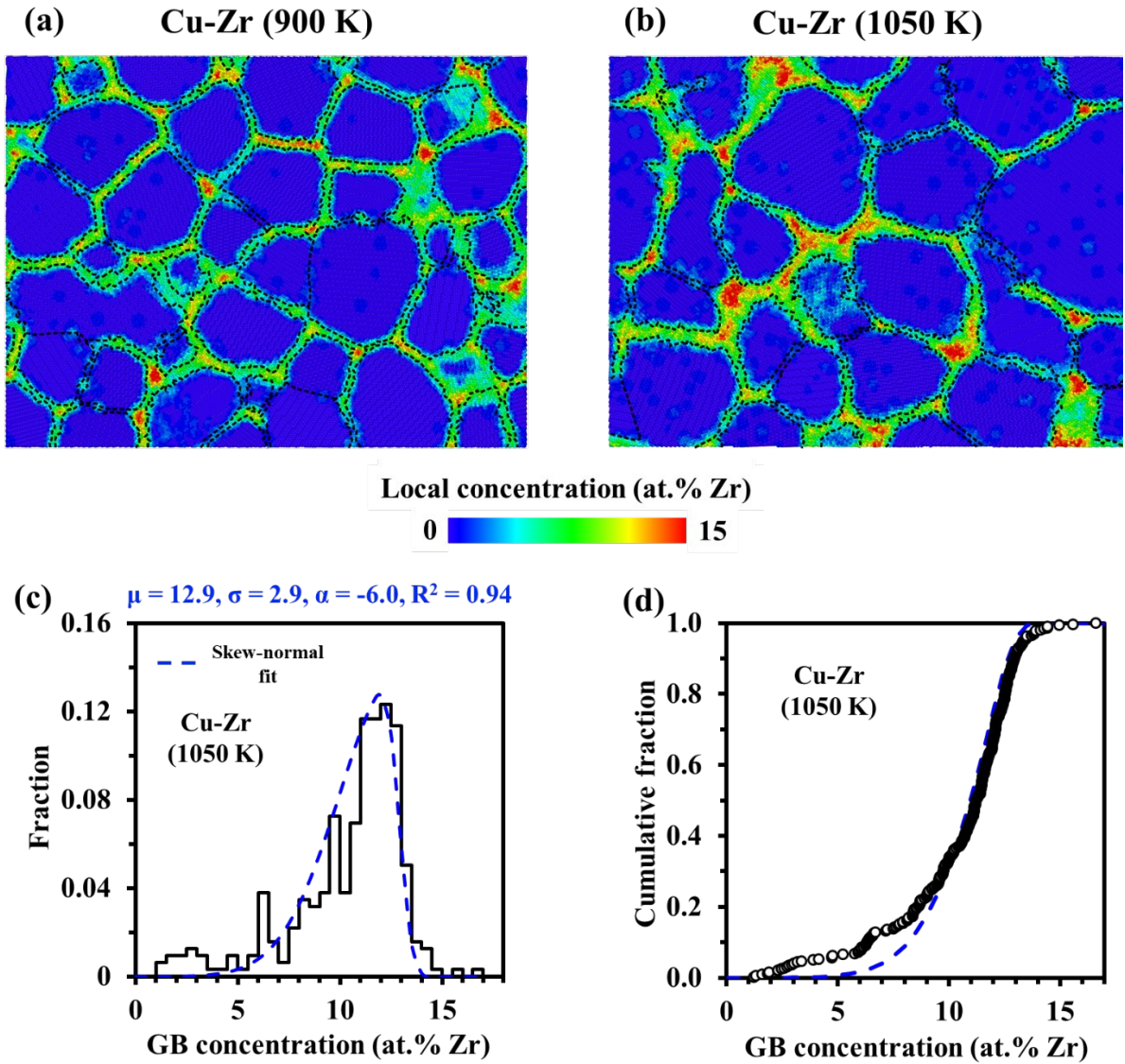

**Figure 7.** Solute distribution within the GB network of the Cu-Zr alloy after equilibrating at (a) 900 K and (b) 1050 K. The GB regions are outlined with dashed lines and local concentration is used to color the image. At 900 K, solute rich regions are smaller in size and distributed throughout the GB network. At 1050 K, the solute rich regions are larger and tend to cluster near one another, while other boundaries appear to be depleted of solute. (c)-(d) Statistical populations of GB concentration for the Cu-Zr polycrystal equilibrated at 1050 K, where depletion is evidenced by the large skew toward low GB concentrations.

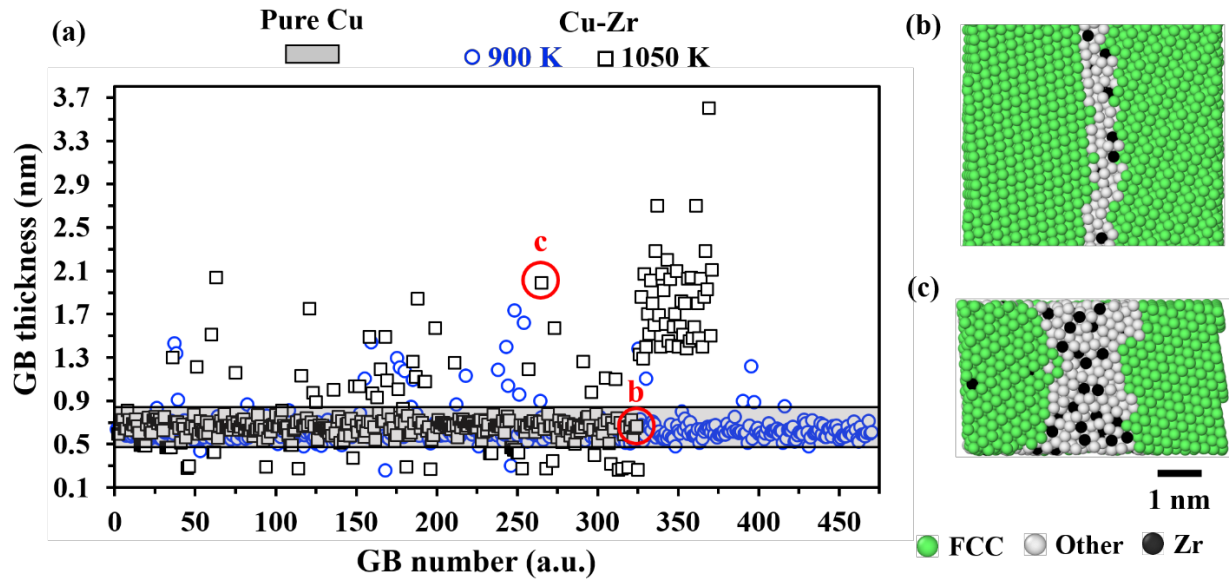

**Figure 8.** (a) Compiled GB thickness measurements for the Cu-Zr alloy polycrystal equilibrated at 900 K (blue circles) and 1050 K (black squares). The grey shaded region corresponds to the GB thickness range measured from the pure Cu polycrystal equilibrated at 900 K, representing a comparison point for a specimen with only ordered GBs. Specific examples of GBs in the Cu-Zr alloy at 1050 K with thicknesses of (b) 0.7 nm and (c) 2 nm thick are shown, with the interface viewed in an edge-on condition. Cu atoms are colored by CNA, with green signaling an FCC structure and white corresponding to GB atoms, while Zr atoms appear black. The sample equilibrated at higher temperature is found to have more boundaries that lie outside the pure Cu reference range, which are found to be amorphous complexions.

**Twin boundaries in Cu-Zr (1050 K)**

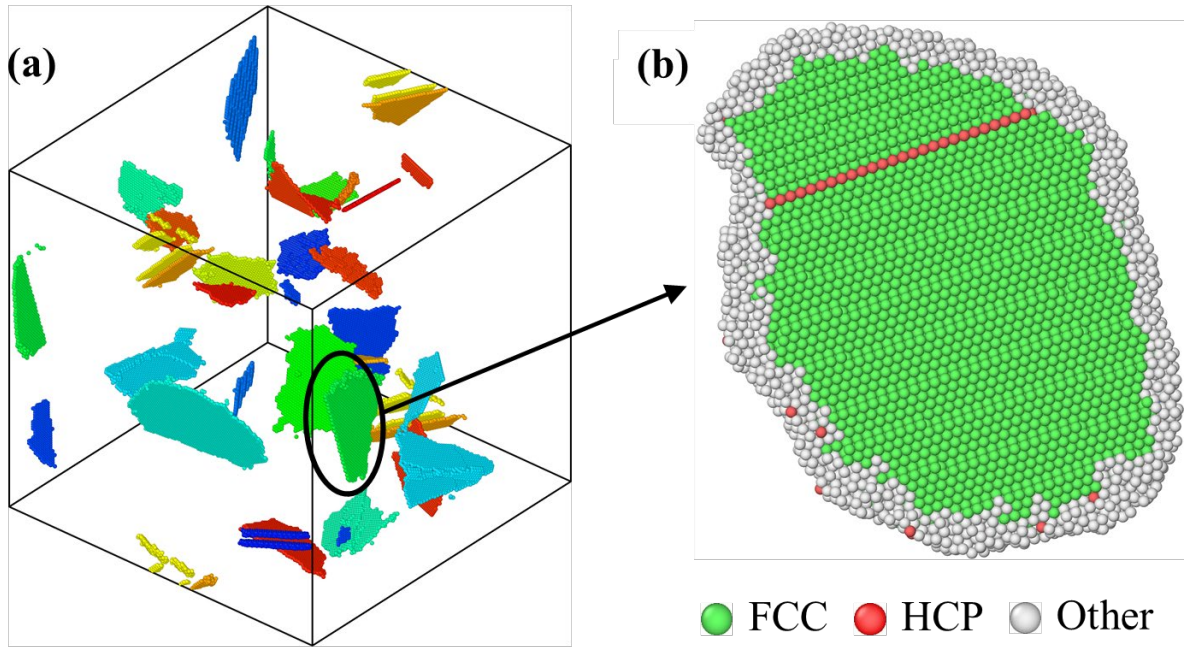

**Figure 9.** (a) GBs with thickness less than 0.45 nm in the Cu-Zr alloy equilibrated at 1050 K, with the atoms colored according to the GB number. (b) Atomic structure analysis using CNA shows that these flat, planar interfaces are in fact intragranular annealing twins, not traditional GBs. The twin boundaries have low solute concentrations of less than 0.1 at.% Zr, matching the low concentration observed in the grain interiors, and appear as a single plane of hcp atoms.

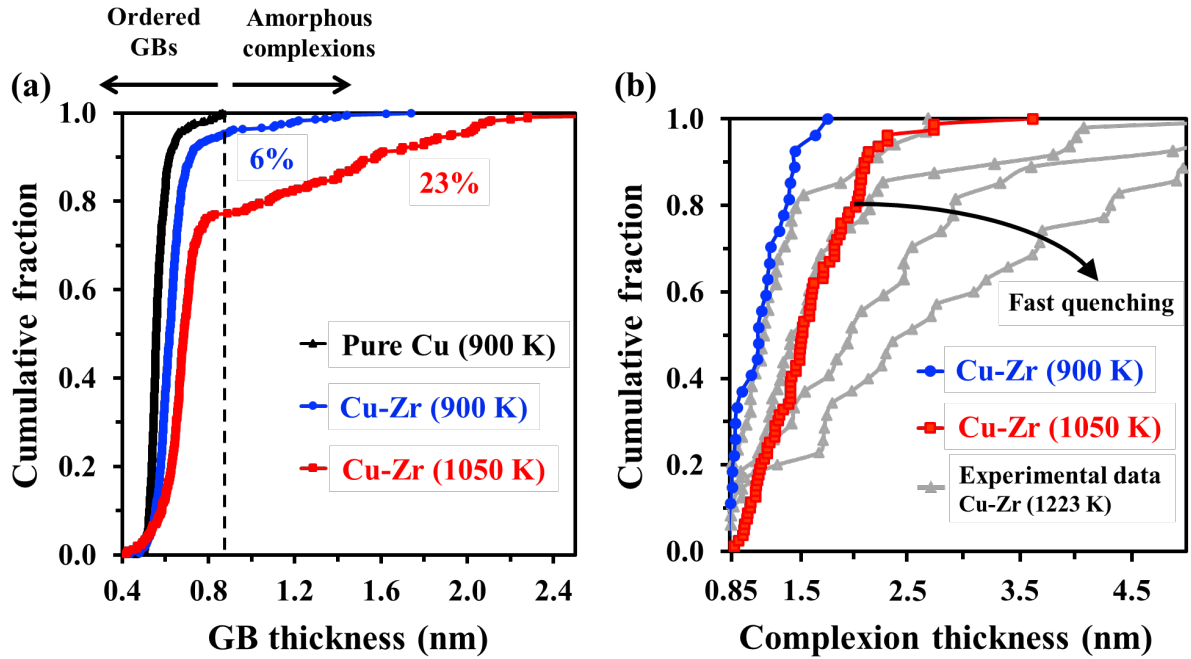

**Figure 10.** (a) Cumulative distribution functions of GB thickness for pure Cu at 900 K, the Cu-Zr alloy at 900 K, and the Cu-Zr alloy at 1050 K. The shift in the curves to the right upon alloying and then increasing temperature shows that thicker GB complexions are found within the polycrystalline GB network. The fact that a larger fraction of boundaries lie to the right of the pure Cu reference state also shows that higher temperatures lead to more interfaces transforming to amorphous complexions. (b) The GB thickness data from the alloyed samples is shown with the thinnest boundaries corresponding to the ordered pure Cu reference state truncated from each distribution. These complexion thickness distributions can be compared to experimental literature reports of amorphous complexion thickness, such as the example shown from Ref. [79] for Cu-3 at.% Zr and Cu-5 at.% Zr annealed at 1223 K (950 °C) and then quenched at different rates.

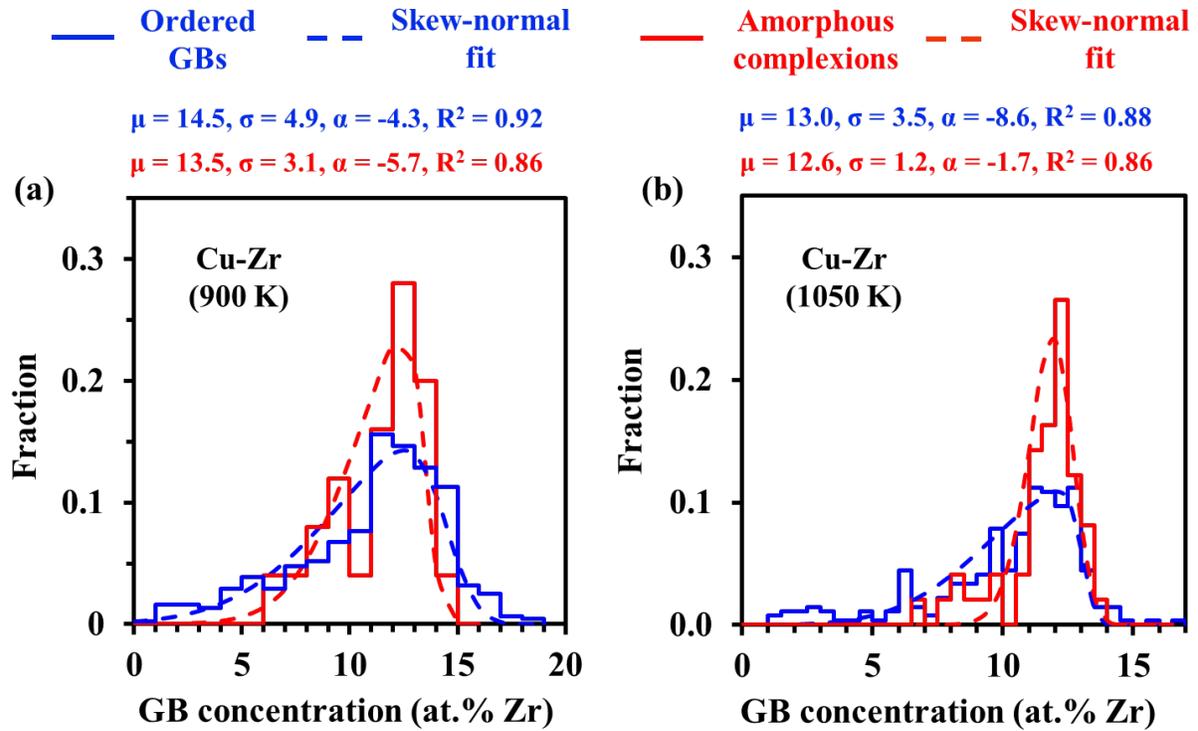

**Figure 11.** GB concentration data for the Cu-Zr polycrystal equilibrated at (a) 900 K and (b) 1050 K, separated into different populations for ordered GBs and amorphous complexions. The existence of two sub-populations of GB concentration for a given polycrystalline sample, each associated with one complexion type, provides further evidence of the co-existence of two complexion types with the GB network.

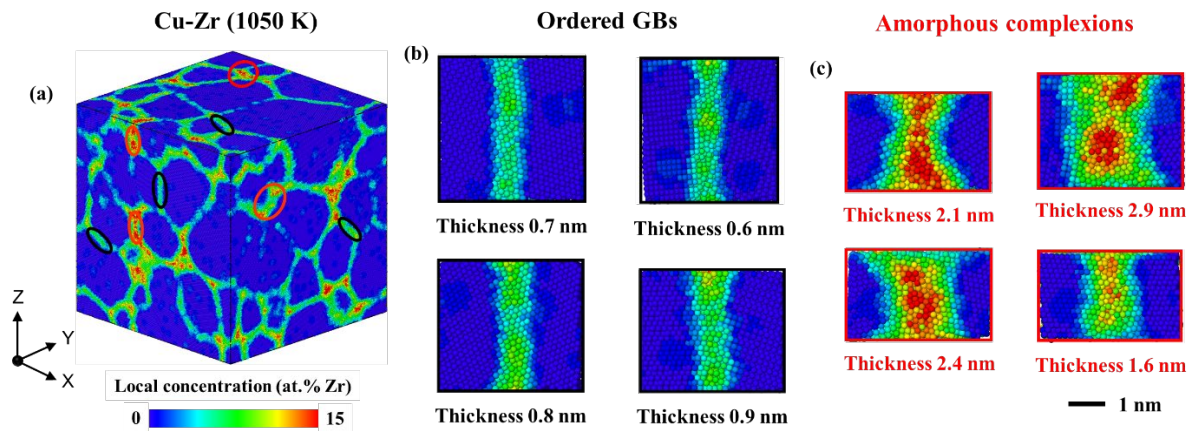

**Figure 12.** Examples of doped ordered GBs (denoted by black circles) and thick amorphous GB complexions (denoted by red circles) in the Cu-Zr polycrystal at 1050 K, with atoms colored according to local Zr concentration. Inspection of specific (b) ordered GBs and (c) amorphous complexions shows enrichment in Zr for all as compared to the grain interior, but also noticeable chemical heterogeneity within a given amorphous complexion.

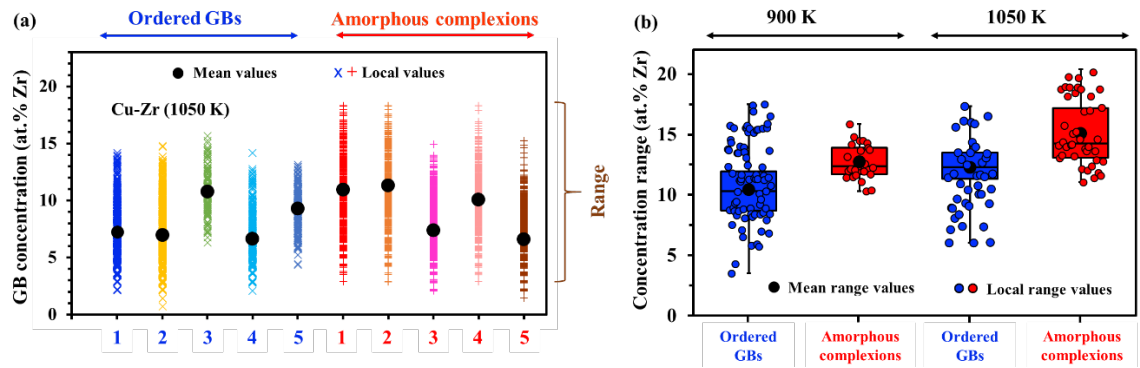

**Figure 13.** (a) Mean and local/atomic measurements of solute distribution at example ordered GBs and amorphous complexions in the Cu-Zr alloy equilibrated 1050 K. (b) Compiled measurements of the local/atomic concentration ranges found at specific boundaries in Cu-Zr alloys equilibrated at 900 K and 1050 K.